\documentclass [12pt,a4paper]{article}
\newcommand{\be}{\begin{equation}} \newcommand{\ee}{\end{equation}}
\newcommand{\ba}{\begin{eqnarray}} \newcommand{\ea}{\end{eqnarray}}
\newcommand{\bes}{\begin{equation*}} \newcommand{\ees}{\end{equation*}}
\newcommand{\bas}{\begin{eqnarray*}} \newcommand{\eas}{\end{eqnarray*}}
\newcommand{\nn}{\nonumber}
\newcommand{\mc}{\mathcal}

\begin{document}
\centerline{\bf\Large Cosmological Implications of a Scale Invariant}
\centerline{\bf\Large Standard  Model}

\bigskip
\centerline{\bf \large Pankaj Jain, Subhadip Mitra and Naveen K. Singh}

\bigskip
\centerline{Physics Department, I.I.T. Kanpur, India 208016}

\bigskip
\begin{center}
\begin{minipage}{0.9\linewidth}
{\small {\bf Abstract:}
We generalize the standard model of particle physics such it displays
global scale
invariance. The gravitational action is also suitably modified such that
it respects this symmetry. This model is interesting since the
cosmological constant term is absent in the action.
We find that the scale symmetry is broken
by the recently introduced cosmological symmetry breaking mechanism.
This simultaneously generates all the dimensionful parameters such as the
Newton's gravitational constant, the particle masses and the vacuum or dark
energy. We find that in its simplest version the
model predicts the Higgs mass to be very small,
which is ruled out experimentally. We further generalize the model such
that it displays local scale invariance. In this case the Higgs particle
disappears from the particle spectrum and instead we find a very massive
vector boson. Hence the model gives a consistent description of particle
physics phenomenology as well as fits the cosmological dark energy.}\end{minipage}\end{center}

\section{Introduction}
Explaining the smallness of the observed cosmological constant is a fundamental problem
in physics. The problem is acute since particle physics models suggest a value for
the vacuum energy, which essentially acts as a cosmological constant, many orders of
magnitude larger then the current bound \cite{Weinberg,Peebles,Padmanabhan,Copeland,Carroll,Sahni,Ellis}. A possible approach to solving this problem has been
suggested by Jain and Mitra \cite{Jain}. They argue that it may be possible to impose a symmetry,
called pseudo-scale invariance, on the model of fundamental physics. This symmetry is
related to the standard scale symmetry but is not identical \cite{Cheng,Cheng1}.
In particular
Jain and Mitra \cite{Jain} show that this symmetry is non anomalous. Hence as long as the symmetry
is not broken by some mechanism, the cosmological constant is predicted to be exactly
zero both in classical as well as quantum theory. It is also not possible to break
pseudo-scale invariance spontaneously since the symmetry does not permit any mass
term in the action. Hence in this case the ground state of the theory is necessarily such that
$<0|\phi|0>=0$, where $\phi$ is any field introduced in the action.

Jain and Mitra \cite{Jain} also introduced a new idea for breaking fundamental symmetries. They
argue that if we introduce a scalar field with sufficiently shallow potential then we can
get a classical space-independent solution to the equations of motion, where
the field has a very slow time dependence and is non-zero over the lifetime of
the universe. This is in direct analogy to the slow roll models of dark energy
\cite{Wetterich,Ratra,Fujii,Chiba,Carroll1,Caldwell,Uzan,Amendola,Bean}.
The quantum solution is then constructed by making a quantum expansion around a
time-dependent classical solution. The global minimum of the potential is irrelavant
since the field never reaches this value during its cosmological evolution. The time-dependent
classical solution breaks fundamental symmetries, including pseudo-scale invariance,
and hence yields a non-zero value of dark energy. In the adiabatic limit this
essentially acts as a cosmological constant.

If we impose pseudo-scale invariance, the standard higgs mechanism is not applicable
for spontaneous breakdown of the standard model of particle physics. This is because
we are not allowed to introduce any mass term in the action. Jain and Mitra
\cite{Jain} suggest that
the standard model $SU(2)_L\times U(1)$ may also be broken cosmologically.
In the present paper we construct an explicit model, which implements
cosmologically broken pseudo-scale invariance.

\section{Pseudo-scale invariance}
We first review the idea of pseudo-scale invariance \cite{Cheng,Jain}.
As shown in Ref. \cite{Cheng} the standard scale transformations can be split
into a general coordinate transformation and pseudo-scale transformation.
Let $\Phi(x)$ represent a scalar field, $A_\mu(x)$ a vector field, $\Psi(x)$
a spin half field and $g^{\mu\nu}(x)$ the metric. In four dimensions
the pseudo-scale transformation can be written as follows:
\ba
x & \rightarrow & x\, , \cr
\Phi &\rightarrow & \Phi/\Lambda\, ,\cr
g^{\mu\nu} &\rightarrow & g^{\mu\nu}/\Lambda^2\, , \cr
A_\mu &\rightarrow & A_\mu\, ,\cr
\Psi &\rightarrow & \Psi/\Lambda^{3/2}\, .
\label{eq:PSU}
\ea
As shown in Ref. \cite{Cheng}, the standard kinetic energy terms for scalar, spin half
fermion and vector particles are invariant under this transformation. All the
terms which involve no parameters with dimensions of mass are found to be invariant.
The gravitational action is modified such that \cite{Cheng,Cheng1},
 \begin{equation}
 {1\over 4\pi G} R \rightarrow \beta \Phi^*\Phi R
 \end{equation}
 where $G$ is the gravitational constant, $R$ the Ricci scalar and
 $\beta$ a dimensionless constant.
The modified action involves no mass parameter and is invariant under the
pseudo-scale transformation.

We next regulate the matter part of the action using dimensional regularization.
We demonstrate that we can consistently regulate the action, such that the
regulated action is invariant under a transformation in $n$-dimensions which
reduces to pseudo-scale invariance in 4-dimensions. In Ref. \cite{Jain}
we have already demonstrated
this for scalar fields. The action for complex scalar fields in
$n$-dimensions can be written as
\begin{eqnarray}
\mathcal{S}=\int d^nx\sqrt{-g}\left[g^{\alpha\beta}
\partial_{\alpha}\Phi^{*}(x)\partial_{\beta} \Phi(x) -
\lambda (\sqrt{-g})^{(n-4)/n} (\Phi^{*}(x)\Phi(x))^2\right].
\label{eq:ActionPhi}
\end{eqnarray}
The action is invariant under the transformation
\ba
x & \rightarrow & x\, , \cr
\Phi &\rightarrow & \Phi/\Lambda^{a(n)}\, ,\cr
g^{\mu\nu} &\rightarrow & g^{\mu\nu}/\Lambda^{b(n)}
\label{pseudo}
\ea
where $b(n) = 4a(n)/(n-2)$ and we may choose $a(n)$ to be any function of
$n$.
We may now generalize this to gauge fields. We consider a U(1) gauge theory.
Its action in $n$ dimensions can be written as
\be
\mathcal{S}
= \int d^nx \sqrt{-g}\Bigg(g^{\mu\nu}(D_\mu\Phi)^*(D_\nu\Phi)
-(\sqrt{-g})^{(n-4)/n}\lambda(\Phi^*\Phi)^2 - \frac14
(\sqrt{-g})^{(4-n)/n}  g^{\mu\nu}g^{\alpha\beta}\mathcal{F}_{\mu\alpha}
\mathcal{F}_{\nu\beta}\Bigg)
\ee
where $D_\mu = \partial_\mu - igA_\mu$ is the gauge covariant derivative
and ${F}_{\mu\nu} = \partial_\mu A_\nu - \partial_\nu A_\mu$ is the field
strength tensor. This action is invariant under the transformation law shown
in Eq. \ref{pseudo} with $A_\mu\rightarrow A_\mu$. The pseudo-scale invariant
action for fermion fields can also be generalized to $n$ dimensions.
The action using the tetrad formalism can be written as
\ba
\mathcal{S} =  \int d^nx \sqrt{-g} \Bigg[\bar{\Psi}\gamma^ce^\mu_ci
\Bigg(D_\mu-\frac12\sigma_{ab}e^{b\nu}
(D_\mu e^a_\nu - \Gamma^\rho_{\mu\nu}e^a_\rho)\Bigg)\Psi\Bigg]
\label{eq_S_fermion}
\ea
where $e^\mu_a$ is the tetrad. The transformation rule for the fermion fields
under pseudo-scale transformation in $n$ dimensions is given by
\be \Psi\rightarrow \Psi/\Lambda^{c(n)} \ee
where $c(n) = a(n) (n-1)/(n-2)$.

The above discussion shows that we can impose exact pseudo-scale invariance
on the regulated action which consists of scalar, spin half fermion and
gauge fields. Hence it is clear that we can consistently impose pseudo-scale
invariance on the Standard model of particle physics. However the main
problem with the action above is that it is not invariant under general
coordinate transformation. This is not a problem as long as treat gravity
classically but is an obstacle to quantization of gravity.
This appears to be necessary as long as we are
making a perturbative expansion around the minimum of the potential, $\Phi=0$.
However it appears that this need not be true if the theory is cosmologically
broken. In this case we make a quantum expansion around a classical time
dependent solution $\Phi_{\rm cl}(t)$. The mass scale of the classical solution
$|\Phi_{\rm cl}(t)|$ itself provides us with a scale necessary 
for regulating the
action.
Hence we may regulate the scalar field action, Eq. \ref{eq:ActionPhi}, in
$n$ dimensions by multiplying the $(\Phi^*\Phi)^2$ term by
$(\Phi^*\Phi)^{-\epsilon}$, where $\epsilon=(n-4)/(n-2)$.
The theory now displays both pseudo-scale and general coordinate invariance.
We may expand the field $\Phi(x) = \Phi_{\rm cl}(t) +
\hat \Phi(x)$, where $\hat \Phi$ represent the quantum fluctuations.
Expanding $(\Phi^*\Phi)^{-\epsilon}$ in powers of the quantum fluctuations
$\hat \Phi$ we find that the leading term is proportional to 
$(\Phi_{\rm cl}^*\Phi_{\rm cl})^{-\epsilon}$ and well defined. It remains
to check the contributions due to terms involving higher powers in $\epsilon$,
which we do not address in this paper.
We also illustrate this by using nonlocal regularization
\cite{Kleppe,Kleppe1}. We use the smearing function,
\ba
f() = f\left({1\over\xi^2\Phi^{*}\Phi}D^2\right)
\label{eq:nonlocal}
\ea
where $\xi$ is a dimensionless
parameter and $D_\mu$ is a covariant derivative. The operator $D^2=g^{\mu\nu}
D_\mu D_\nu$. 
The function $f()$ is chosen such that it regulates the
loop integrals at mass scale
$\xi|\Phi_{\rm cl}|$.
A suitable choice for $f()$ is the exponential function used in Ref.
\cite{Kleppe,Kleppe1}.
It is clear that the smearing function is invariant under both scale and
general coordinate transformation. Hence this allows us to regulate the
theory while preserving scale and general coordinate invariance.
This regulator is available to us as long as the scale invariance is
broken cosmologically. An explicit demonstration that this indeed defines a
consistent quantum field theory is postponed to future research.

\section{Standard Model}
In this section we construct pseudo-scale invariant Standard model of particle
physics.
We point out that the only terms in the Standard model
which violate the pseudo-scale invariance are the Higgs boson mass terms. Such
terms are required in order that the Standard model gauge invariance
may be broken spontaneously. In the present case we break the gauge
invariance cosmologically due to the slowly varying background scalar field.
Hence we can set the mass terms to zero and construct a model with pseudo-scale
invariance.

We introduce the complex doublet Higgs field,
\[ \mc H(x) = \left( \begin{array}{c}
h_1(x) \\
h_2(x) \end{array} \right) .\]
 We assign $\mc H$ a charge $+1/2$ under $U(1)$ so that its
transformation under $SU(2)\times U(1)$ is given by:
\bas
\mc H \rightarrow e^{i(\alpha^a\tau^a+\beta/2)}\mc H
\eas
where $\tau^a = \sigma^a/2$. Using these two fields we can write the
$SU(2)\times U(1)$ invariant action in 4 dimensions,
\ba
\mathcal{S} &=& \int d^4x \sqrt{-g}\Bigg[-{\beta\over 4} \mc H^\dag \mc H R
+g^{\mu\nu} (D_\mu \mc H)^\dag(D_\nu \mc H) - \frac14
 g^{\mu\nu}g^{\alpha\beta}(\mathcal{A}_{\mu\alpha} \mathcal{A}_{\nu\beta}
\nn\\
 &+& \mathcal{B}_{\mu\alpha} \mathcal{B}_{\nu\beta})
- \lambda    (\mc H^\dag \mc H)^2\Bigg]
\label{eq:S_EW}
\ea
where $D_\mu = \partial_\mu -ig{\bf A}_\mu - ig'B_\mu$,
$B_\mu$ is the $U(1)$ gauge field, ${\bf A}_\mu = \tau^aA^a_\mu$
is the $SU(2)$ gauge field multiplet and $\mathcal{A}_{\mu\nu}$ and
$\mathcal{B}_{\mu\nu}$ are the respective field strength tensors for the
gauge fields $A_\mu$ and $B_\mu$.
Here we have displayed the action only in four dimensions. The action can
be suitably regulated as discussed in section 2. We can also include fermions
as discussed in section 2. We emphasize that since we have imposed exact
pseudo-scale invariance we are not allowed to introduce the cosmological
constant term in the action. Cosmological constant can also not be generated
by quantum fluctuations since the pseudo-scale invariance is preserved by
the regulated action. Hence this solves the standard cosmological constant
problem. Alternative  approaches to address the
cosmological constant problem are discussed in Ref.
\cite{Weinberg,Aurilia,VanDer,Henneaux,Brown,Buchmuller,Henneaux89,Sorkin,Sundrum}.

\section{Classical Solution}
We first consider a real scalar field $\Phi(x)$ coupled to gravity.
We later generalize this to the case where the scalar field is identified as
the Higgs multiplet.
The action can be written as:
\ba
\mc{S} = \int d^4x \sqrt{-g}\left[\frac12g^{\mu\nu}\partial_\mu\Phi\partial_\nu\Phi
-\frac{\lambda}{4}\Phi^4 -\frac{\beta}{8}\Phi^2R \right]
\ea
where the gravitational action has been modified by the relation:
$R/(2\pi G) \rightarrow \beta \Phi^2R$, to maintain the pseudo-scale
invariance \cite{Cheng,Cheng1}. This model falls in the general category
of scalar-tensor models whose cosmological implications
have been studied extensively in recent literature
\cite{Agarwal,Barrow,Barenboim}.
We take the metric as FRW metric with $k=0$, {\textit i.e.}, the non-zero components of
the metric are given by $g_{00} = 1$, $g_{11}=g_{22}=g_{33} = -a^2(t)$,
where $a(t)$ is the scale factor. If we neglect the
quantum fluctuations, then $\Phi(x) \approx \eta(t)$, where $\eta(t)$ is the
space-independent classical solution.
The Einstein equation generalizes such that,
\ba
R_{\mu\nu} - \frac12Rg_{\mu\nu} &=& \frac{4}{\beta\eta^2}T_{\mu\nu}.\label{eq:eq_gravity}
\ea
Here $T_{\mu\nu}$, the stress energy tensor, is given by:
$T_{00} = \dot\eta^2/2 +\lambda\eta^4/4$,
$T_{ii} = -g_{ii}( \dot\eta^2/2 -\lambda\eta^4/4)$,
with zero off-diagonal components.
The $0-0$ component of Eq. \ref{eq:eq_gravity} gives:
\bas
H_0^2 = \left(\frac{\dot a}{a}\right)^2 = \frac{1}{3\beta}
\left[2\left(\frac{\dot\eta}{\eta}\right)^2 + \lambda\eta^2\right]
\eas
and the $i-i$ components give:
\bas
2\left(\frac{\ddot a}{a}\right) + \left(\frac{\dot a}{a}\right)^2 = - \frac1\beta
\left[2\left(\frac{\dot\eta}{\eta}\right)^2 - \lambda\eta^2\right].
\eas
The equation of motion for $\eta$ is found to be,
\ba
\ddot\eta + 3H_0\dot\eta + \left(\frac{\beta R}{4} + \lambda\eta^2\right)\eta &=& 0.
\label{eq:eq_eta}
\ea

For a vacuum energy dominated universe, evolution of the scale factor is given by $a(t) = a_0e^{H_0t}$ where $H_0$ is
a constant. So the Ricci scalar becomes $R = -12 H_0^2$. In this case, we see that, Eq. \ref{eq:eq_eta}
allows a constant solution \cite{Jain07} for $\eta$, given by,
\ba
\eta = \sqrt{\frac{3\beta}{\lambda}}H_0.\label{eq:soln_eta}
\ea
If we use this to estimate the vacuum energy density, we get,
\bas
\rho_{V} = \frac14\lambda\eta^4 = \frac34H_0^2(\beta\eta^2)\approx\frac3{8\pi}H_0^2M_{PL}^2
\eas
where in the last step we have put $\beta\eta^2 \approx (2\pi G)^{-1} = M_{PL}^2/2\pi$.

We may now make a quantum expansion around this solution $\Phi(x)=\eta(t)
+ \phi(x)$, where $\phi(x)$ denote the quantum fluctuation. The total
Lagrangian becomes,
\ba
 \mc L = \frac12g^{\mu\nu}\partial_\mu\phi\partial_\nu\phi
-3\beta H_0^2\phi^2 - \sqrt{3\beta\lambda}H_0\phi^3 - \frac14\lambda\phi^4
+\frac3{8\pi}H_0^2M_{PL}^2.
\ea
Hence the quantum field, $\phi$ gains a mass given by, $m_\phi = \sqrt{6\beta}H_0$.

We next interpret the scalar field as the Higgs field responsible for
electroweak symmetry breaking.
The entire calculation may be repeated assuming that the scalar field is the
standard model
Higgs field multiplet $\mc H(x)$.
The calculation is unchanged with $\Phi^2$ replaced by $2|{\mc H}^\dagger
{\mc H}|$.

 The electroweak symmetry is broken with the correct
mass spectrum for the standard model particles if we assume that the classical
solution $\eta^2=|{\mc H}^\dagger {\mc H}|=M^2_{\rm EW}$, where
$M_{\rm EW}$ is the
electroweak scale. We predict the mass of the Higgs field of order
$\sqrt{\beta} H_0$ which turns out to be of order $H_0 M_{\rm PL}/M_{\rm EW}$.
With the current value of the Hubble constant, this turns out to be very
small. Hence this model is ruled out. It would be interesting to investigate
if it is possible to enhance the scalar sector in order to avoid this
conclusion. Alternatively we may interpret the scalar field as some other
multiplet responsible for breakdown of a grand unified gauge group. In
any case we do not pursue this further in the present paper. Instead
we consider the possibility \cite{Cheng,Cheng1} of local scale invariance in the
next section.

\section{Local Pseudo-scale Invariance}

We next gauge the pseudo-scale invariance by introducing the Weyl vector meson
$S_\mu$ following the approach Ref. \cite{Cheng,Cheng1}. Local scale invariance
has also been considered in Ref. \cite{Padmanabhan85,Hochberg,Wood,Wheeler,Feoli,Pawlowski,Nishino,Demir}. The local scale invariance
may also be broken cosmologically if the classical solution
for the scalar field has non-zero value. We expect that in this case the Weyl
meson would become massive and one of the scalar fields will be
eliminated from the particle spectrum. We first consider a real scalar field
and later replace this field by the standard model Higgs multiplet.

We demand the action to be invariant under local pseudo-scale transformation such that in Eq.
\ref{eq:PSU} $\Lambda$ becomes a
function of $x$, i.e. $\Lambda = \Lambda(x)$. To compensate for the $x$ dependence
of $\Lambda$ we need to introduce an additional vector field $S_\mu$ and replace
$\partial_\mu$ by $(\partial_\mu - fS_\mu)$ in the Lagrangian. The Lagrangian
becomes,
\ba
\mc L = -{\beta\over 8} \Phi^2 \tilde R+ \mc L_{matter}
\ea
where
\ba
\mc L_{matter} = \frac12g^{\mu\nu}(\mc D'_\mu\Phi)(\mc D'_\nu\Phi) - 
{\lambda\over 4}\Phi^4 -
{1\over 4}g^{\mu\rho}g^{\nu\sigma}E_{\mu\nu}E_{\rho\sigma}\ ,
\ea
$f$ is the gauge coupling constant, $E_{\mu\nu}=\partial_\mu S_\nu-
\partial_\nu S_\mu$, $\tilde R$ is a modified curvature scalar
\cite{Cheng,Cheng1},
 invariant under local pseudo-scale transformation, and
\ba
\mc D'_\mu &=& \partial_\mu - fS_\mu
\ea
is the gauge covariant derivative.
Under pseudo-scale transformation the vector field $S_\mu$ transforms as
\ba
S_\mu \rightarrow S_\mu - \frac 1f\partial_\mu ln(\Lambda(x))\,.
\ea
The scalar $\tilde R$ is found to be
\ba
\tilde R = R - 6f S^\kappa_{\ ;\kappa} - 6f^2 S\cdot S
\ea
where $R$ is the standard curvature scalar.

We next obtain the classical equations of motion for this locally scale
invariant Lagrangian. The Einstein equation gets modified to
\ba
\Phi^2 B^{\alpha\beta} + \partial_\lambda(\Phi^2) C^{\lambda \alpha\beta}
+(\Phi^2)_{;\lambda;\kappa} D^{\alpha\beta \kappa\lambda} ={4\over \beta}
T^{\alpha\beta}
\ea
where $T^{\alpha\beta}$ is the contribution obtained from all terms other than
$\tilde R $ in the Lagrangian. The tensors $B_{\alpha\beta}$, $C^{\lambda}_{\ \alpha\beta}$
and $D_{\alpha\beta}^{\ \ \ \kappa\lambda}$ are given by,
\bas
B_{\alpha\beta} &=& -{1\over 2} g_{\alpha\beta} R + R_{\alpha\beta} +
3f^2g_{\alpha\beta} S\cdot S - 6f^2 S_\alpha S_\beta\, ,\\
C^{\lambda}_{\ \alpha\beta} &=& -3 f g_{\alpha\beta}S^\lambda +3f\left(S_\beta
g^\lambda_\alpha +S_\alpha g_\beta^\lambda\right)\, ,\\
D_{\alpha\beta}^{\ \ \ \kappa\lambda} &=& -{1\over 2}\left(g^\lambda_\alpha
g_\beta^\kappa  + g_\alpha^\kappa g^\lambda_\beta \right)
+ g_{\alpha\beta}g^{\lambda\kappa}\, .
\eas
The energy momentum tensor, including only the field $\Phi$ and $S_\mu$, may
be written as,
\ba
T_{\mu\nu} = -{\mc L}_{matter}g_{\mu\nu} + \mc D'_\mu \Phi \mc D'_\nu \Phi
- {1\over 2}\left(E_{\alpha\nu}E_{\beta\mu}g^{\alpha\beta}
+ E_{\mu\alpha}E_{\nu\beta}g^{\alpha\beta}\right).
\ea
The equations of motion for $\Phi$ and $S_\mu$ fields are given by,
\bas
g^{\mu\nu}\partial_\nu (\partial_\mu\Phi-f S_\mu\Phi)+
{\beta\over 4} \Phi\tilde R+(\partial_\mu\phi - fS_\mu\Phi)\left[{1\over 2}
g^{\mu\nu}g^{\alpha\beta}\partial_\nu g_{\alpha\beta} + \partial_\nu
g^{\mu\nu}\right]\\
+ fg^{\mu\nu} S_\nu\left(\partial_\mu\Phi - fS_\mu\Phi\right) + \lambda \Phi^3
=0
\eas
and
\bas
\partial_\nu\left[g^{\nu\rho}g^{\mu\sigma}(\partial_\rho S_\sigma
-\partial_\sigma S_\rho)\right] + {1\over 2} g^{\nu\rho} g^{\mu\sigma}
(\partial_\rho S_\sigma - \partial_\sigma S_\rho)g^{\alpha\beta}\partial_\nu
g_{\alpha\beta} + {3\over 2}\beta f^2\Phi^2 g^{\eta\mu}
S_\eta\\
-fg^{\mu\nu} \Phi \mc D'_\nu\Phi - {3\over 4}f\beta g^{\mu\kappa} \partial_\kappa \Phi^2
= 0
\eas
respectively.

The equations simplify considerably if we drop the space derivatives. The
time-time component of the Einstein equation gives
\ba
\Phi^2\left(R_{00} -{R\over 2}\right) + 3\Phi^2f^2(S^iS_i - S_0S_0)
+ 3f S_0 \partial_0(\Phi^2) - g^{ij}\partial_0(\Phi^2) \Gamma^0_{ij} =
{4\over \beta} T_{00}\,.
\ea
The space-space component of the Einstein equation gives
\ba
\Phi^2\left(R_{ij} - {1\over 2} g_{ij} R\right)
+ 3f^2\Phi^2 g_{ij} S\cdot S - 6f^2\Phi^2 S_iS_j
- 3fg_{ij} S^0\partial_0\Phi^2
+ g_{ij} \partial_0\partial_0\Phi^2
= {4\over\beta} T_{ij}\,.
\ea
The equation of motion for the scalar field reduces to
\ba
\ddot \Phi - f\Phi \dot S_0 + 3\dot \Phi{\dot a\over a} - f^2S_0^2\Phi
+\lambda\Phi^3 - 3f\Phi S_0{\dot a\over a} + {\beta\over 4} \Phi\tilde R = 0
\ea
where the dots represent derivatives with respect to time.
We also display the equations for $S_0$ and $S_i$.
\ba
f S_0 = {\dot\Phi\over \Phi}\,,
\label{eq:S0}
\ea
\ba
\ddot S_i + {\dot a\over a}\dot S_i + f^2\Phi^2 S_i + {3\over 2}\beta f^2\Phi^2 S_i = 0\,.
\label{eq:Si}
\ea
It is clear from Eq. \ref{eq:S0} that as long as $\Phi$ is time independent
$S_0=0$. Similarly from Eq. \ref{eq:Si} we find that $S_i$ has a very
large mass. The mass is of order $fM_{Pl}$, which is huge as long as $f$ is
not too small \cite{Cheng}. Hence $S_i$ would reach its minimum very early
during the cosmological evolution
and it is reasonable to set $S_i=0$. This implies that the classical
solution remains the same as that obtained for the case of global scale
invariance if we assume $\dot\Phi=0$. However now the major advantage is that the very light scalar
field has disappeared from the particle spectrum. We see this easily by making
a gauge transformation such that
\ba
\Phi \rightarrow {\Phi \over 1+\phi(x)/\eta(t)}
\ea
with the corresponding transformation of other fields. The action is invariant
under this transformation.
The field $\phi$ has disappeared from the spectrum,
essentially becoming the longitudinal mode of $S_\mu$.

We next generalize the standard model such that it displays local pseudo-scale
invariance. This can be done by replacing $\mc D_\mu$ in Eq. \ref{eq:S_EW} by
\bas
\mc D''_\mu &=& \mc D_\mu - fS_\mu.
\eas
We again obtain
a classical solution similar to that obtained earlier in this section with
the scalar field $\Phi$ replaced by $2\mc H^\dagger \mc H$.
Hence we find that the scale invariant standard model of
particle physics displays dark energy. It also predicts absence of Higgs
particle. Instead we find a very massive Weyl vector boson.

So far the
solution we have considered does not display cold dark matter. This problem
is best addressed after taking into account the visible matter. The Einstein's
equations are modified in our model since we have imposed scale invariance.
Hence the matter content need not be identical to what is expected in
the standard big bang model. In this case we no longer expect the scalar
field to be time independent classically. Furthermore the Weyl vector field
is also likely to be non-zero. We postpone
this interesting issue to future research.

\bigskip
\noindent
{\bf Acknowledgements:} PJ thanks Ashoke Sen for a useful discussion.

\end{document}